\documentclass[twocolumn,aps,showpacs,longtable]{revtex4}

\usepackage{amsfonts,amsmath,graphicx,latexsym}
\usepackage{epsfig}
\usepackage{psfrag}

\newcommand{\vc}[1]{\mathbf{#1}}

\begin{document}
\title{Subsonic critical velocity at finite temperature}
\author{Patrick Navez, Robert Graham }
\affiliation{
Universitaet Duisburg-Essen,
Universitaetsstrasse 5,
45117 Essen,
Germany}
 

\date{\today}
\begin{abstract}
Based on the dielectric formalism in the generalized 
random phase approximation, 
we generalize the description of a Bose condensed gas 
to allow for a relative velocity  between
the superfluid and normal fluid. In this model, we determine 
the critical velocity  dynamically 
as the transition point between stable and unstable dynamics.
Unlike the zero temperature case, 
at finite temperature the relative critical velocity 
of a dilute Bose gas is lower than  
the sound velocity.  This result illustrates  
one relevant difference that exists 
between a conserving and gapless approximation and other approaches.
\end{abstract}

\maketitle


One of the most fascinating aspects of the observed 
Bose condensation phenomenon is the superfluidity 
property i.e. the ability of the superfluid fraction of 
the Bose gas to move without apparent friction. No dissipation 
occurs provided that the relative motion of the superfluid 
with an obstacle, like an impurity or a local potential or a wall, 
does not exceed the sound velocity. 
Otherwise, the Landau criterion predicts that a phonon can 
be emitted decreasing the velocity of the superfluid \cite{textbook}. 
The 
situation is drastically different if the obstacle corresponds 
to the normal component of the gas. In principle, the Landau 
criterion cannot be applied to the normal fluid as a whole but 
to each of its individual thermal excitation. Unfortunately, some of 
them are highly energetic and display a relative velocity higher 
than the sound velocity. In order to explain the  
persistence of the relative motion between the normal and superfluid, a full 
microscopic description is necessary. A quantum kinetic 
equation explaining this persistence has been proposed in 
\cite{Kirkpatrick} and 
has been subsequently extended in 
\cite{Zaremba,Pomeau,Imamovic}. At low 
temperature i.e. $k_BT < gn$ (where $g=4\pi a/m$, $a$ the scattering length and 
$n$ the total gas density), this equation is an extension of the Beliaev 
approach for a non equilibrium Bose gas and predicts that 
the critical and sound velocity are identical. However, the extension 
of the Beliaev theory including the presence of a thermal component is 
not conserving \cite{HM}. In particular the conservation law for the 
total mass density is missing in the systematic calculation made 
in \cite{Imamovic} using the Keldysh formalism. On the contrary, 
at high temperatures 
($k_BT > gn$), 
the conservation laws are fulfilled but at the 
price of having a gap in the energy spectrum 
rendering possible 
the frictionless motion \cite{Zaremba,Kirkpatrick,Pomeau}. 
Of course, by the Hugenholtz-Pines theorem, 
a gap is forbidden and cannot 
be used to explain the superfluidity 
phenomenon. As a consequence, a gapless and conserving 
approach 
is needed in order to improve the description 
in a large range of temperature.
Recently, such an approach has been developed in both equilibrium 
\cite{Fliesser}
and in non equilibrium \cite{condenson} 
using the generalized random phase approximation and 
has been successful to explain the low-lying excitation observed 
in a trapped Bose gas \cite{Reidl}. 

In this letter, based on this more coherent approach, we delimitate 
the region for which a relative velocity between the normal 
fluid and the superfluid is possible at finite temperature. 
Unlike for other obstacles, the transition region from metastability 
to instability, due to counterflow between normal and superfluid 
components, occurs for velocity lower than the sound velocity. 
The persistence is explained both by the equilibrium and non equilibrium 
approaches. In equilibrium, we introduce an $\eta$-ensemble for which 
the condensed fraction evolves with a velocity $v_s$. The 
study of the dynamical fluctuations around this equilibrium 
allows to establish a critical value of $v_s$ 
above which these become instable. 
In non equilibrium, a previous work has shown that the 
supression  of the binary collision process between condensed and 
non condensed atoms allows precisely this persistence \cite{condenson}. 
Above 
the critical velocity, this collisionless regime disappears 
leading to a damping of the relative velocity. We will show 
that the 
equilibrium formalism provides identical results to the non 
equilibrium one, in particular for the dispersion relation 
of the collective 
excitations.


To start with, we consider a homogeneous Bose condensed gas populated with
atoms with mass $m$ in a volume $V$. For any momentum $\vc{k}$,
$n_c$ is the density of the condensate and
$n'_{\vc{k}}$ the density of the normal cloud  
in the mode $\vc{k}$. 
We extend the formalism developed
in \cite{Fliesser} to the case where the $N_c=V n_c$ condensate particles
are moving with a momentum $\vc{k_s}$. For that 
purpose, we break the $U(1)$ symmetry introducing the 
$\eta -$ensemble by adding the term proportional to the 
field operator $c_{\vc{k_s}}$ to the total many body 
Hamiltonian $H$. Thus, any thermal averaging is written as:
\begin{eqnarray}
\langle A \rangle={\rm Tr}(A e^{-\beta[H- \eta c_{\vc{k_s}}-
\eta^* c^\dagger_{\vc{k_s}}] })/
{\rm Tr}(e^{-\beta[H- \eta c_{\vc{k_s}}-
\eta^* c^\dagger_{\vc{k_s}}] })
\end{eqnarray}
In this way, the condensate
wave function $\langle c_{\vc{k_s}} \rangle=\sqrt{N_c}$
is adjusted to move with a velocity 
$\vc{v_s}=\vc{k_s}/m$. The field 
operator in the momentum space can be written as 
$c_{\vc{k}}=\delta c_{\vc{k}} + \delta_{\vc{k},\vc{k_s}}\sqrt{N_c}$.

The Green's functions describing the non condensed fraction 
can be written as usual \cite{Griffin}:
\begin{eqnarray}
G_{\alpha\beta}(\vc{k},\omega)=-
\int_0^\beta d\tau e^{i \omega \tau}
\langle T_{\tau}[\delta c_{\vc{k} \alpha}(\tau)
\delta c^\dagger_{\vc{k} \beta}] \rangle
\end{eqnarray}
where $\delta c_{\vc{k} \alpha}$ is $\delta c_{\vc{k}}$ for 
$\alpha=1$ and $\delta c^\dagger_{2\vc{k_s}-\vc{k}}$ for 
$\alpha=2$.
The only difference with the previous approach 
\cite{Fliesser} is that the propagator line of 
momentum $\vc{k}$ is coupled to the one with 
momentum $2\vc{k_s}-\vc{k}$ (see Fig.1).
The Green's functions satisfy the Beliaev-Dyson relations:
\begin{eqnarray}\label{BD}
G_{\alpha\beta}=
G_{\alpha\beta}^0 +
G_{\alpha\gamma}^0 \Sigma_{\gamma \delta}
G_{\alpha\gamma}
\end{eqnarray}
where $\Sigma_{\gamma \delta}(\vc{k},\omega)$
are the usual self-energy matrix and
where the Green's functions of the non
interacting system in presence of the momentum
$\vc{k_s}$ is given by:
\begin{eqnarray}
{G_{\alpha\beta}^0}^{-1}(\vc{k},\omega)=
\left( \begin{array}{cc}
\omega-\epsilon_\vc{k}+\mu & 0 \\
0 & -\omega-\epsilon_{2\vc{k_s}-\vc{k}}+\mu
\end{array} \right)
\end{eqnarray}
where
$\epsilon_\vc{k}=\vc{k}^2/(2m)$. Solving 
(\ref{BD}), we obtain:
\begin{eqnarray}
\lefteqn{G_{\alpha\beta}(\vc{k},\omega)=}
\nonumber \\
&\frac{1}{D}
\left( \begin{array}{cc}
{\omega+\epsilon_{2\vc{k_s}-\vc{k}}-\mu +\Sigma_{22}} 
& \Sigma_{12} \\
\Sigma_{21}  & {\epsilon_{\vc{k}}-\omega-\mu +\Sigma_{11}}
\end{array} \right)
\end{eqnarray}
where
\begin{eqnarray}
D
&=&(\omega-\vc{q}.\vc{k_s}/m-A_-)^2 
-(\epsilon_{\vc{q}}+\epsilon_{\vc{k_s}}-\mu+A_+)^2
\nonumber \\
&+&\Sigma_{12}\Sigma_{21}
\end{eqnarray}
$A_\pm = (\Sigma_{11}\pm \Sigma_{22})/2$
and $\vc{q}=\vc{k}-\vc{k_s}$.
The susceptibility function is defined as:
\begin{eqnarray}
\chi_{nn}(\vc{q},\omega)=
-
\int_0^\beta d\tau e^{i \omega \tau}
\langle T_{\tau}[\rho_{\vc{q}}(\tau)
\rho^\dagger_{\vc{q}}] \rangle
\end{eqnarray}
where $\rho_{\vc{q}}=\sum_{\vc{k}} 
c^\dagger_{\vc{k}+\vc{q}} c_{\vc{k}}$.
It can be expressed in terms of the proper dielectric
function $\tilde{\chi}_{nn}$ through:
\begin{eqnarray}
\chi_{nn}(\vc{q},\omega)=
{\tilde{\chi}_{nn}(\vc{q},\omega)}
/({1-g\tilde{\chi}_{nn}(\vc{q},\omega)})
\end{eqnarray}
The proper part of a quantity is the sum of only those 
of its diagrams which remain connected after cutting 
a single interaction line.
All the unknown functions can be expressed in terms 
of regular (i.e. proper and one-particle irreducible) 
quantities, denoted by an upper index:
\begin{eqnarray}
\tilde{\chi}_{nn}(\vc{q},\omega)
&=&
\chi_{nn}^{(r)}(\vc{q},\omega)+
\Lambda_\alpha^{(r)}(\vc{q},\omega)
{\tilde G}_{\alpha \beta}(\vc{k},\omega)
\Lambda_\beta^{(r)}(\vc{q},\omega)
\nonumber \\
\\
{\tilde G}_{\alpha\beta}&=&
G_{\alpha\beta}^0 +
G_{\alpha\gamma}^0 \Sigma^{(r)}_{\gamma \delta}
{\tilde G}_{\alpha\gamma}
\\
\Sigma_{\alpha \beta}(\vc{k},\omega)&=&
\Sigma^{(r)}_{\alpha \beta}(\vc{k},\omega)+
\frac{g\Lambda_\alpha^{(r)}(\vc{q},\omega)
\Lambda_\beta^{(r)}(\vc{q},\omega) }{1-g
\chi_{nn}^{(r)}(\vc{q},\omega)}
\end{eqnarray}
The main difference to \cite{Fliesser} is that the susceptibility 
function is expressed in terms of the momentum 
$\vc{q}=\vc{k}-\vc{k_s}$ relative to the condensate momentum while the 
Green's function is expressed in terms of 
the absolute momentum $\vc{k}$. As a consequence, the 
momentum of the incoming and outgoing lines in the vertex 
function must be 
relabelled adequately as shown in Fig.(2).

Once the regular functions are known, any physical 
quantities can be calculated. In the random phase 
approximation, these functions are found to be:
\begin{eqnarray}
\chi_{nn}^{(r)}(\vc{q},\omega)&=&\chi_{nn}^0(\vc{q},\omega)
/(1-g\chi_{nn}^0(\vc{q},\omega))
\\
\Lambda_\alpha^{(r)}(\vc{q},\omega)
&=&\sqrt{N_c}/(1-g\chi_{nn}^0(\vc{q},\omega))
\\
\Sigma^{(r)}_{\alpha \beta}(\vc{k},\omega)&=&
g(2n-n_c)\delta_{\alpha,\beta}
 \nonumber \\ &+&
g^2N_c\chi_{nn}^0(\vc{q},\omega)
/(1-g\chi_{nn}^0(\vc{q},\omega))
\end{eqnarray}
These are expressed in terms of the susceptibility function 
of the non interacting normal fluid:
\begin{eqnarray}
\chi_{nn}^0(\vc{q},\omega)&=&
\sum_{\vc{k}}
\frac{n'_{\vc{k}}-n'_{\vc{k+q}}}{
\omega -
\frac{\vc{k}.\vc{q}}{m}-\frac{\vc{q}^2}{2m}}
\end{eqnarray}
Below the condensation point, the chemical potential 
is determined through the tadpole condition or 
the Hugenholtz-Pines theorem
$\mu=\epsilon_{\vc{k_s}}+g(2n-n_c)$ and contains 
a kinetic energy term of the condensate particle $\epsilon_{\vc{k_s}}$.
The particle momentum distribution for a  normal fluid at rest 
reads in the Hartree-Fock approximation as 
$n'_{\vc{k}}=1/{[e^{\beta(\vc{k}^2/2m + 2gn-\mu)}-1]}$. 
The equation of state is:
\begin{eqnarray}\label{state}
n=n_c+\frac{1}{\lambda_{th}^{3}}
g_{3/2}(e^{\beta (\mu- 2gn)})
\end{eqnarray}
where $g_n(x)=\sum_{j=1}^\infty x^j/j^n$
and $\lambda_{th}=\sqrt{2\pi /mk_B T}$ denotes the thermal
wavelength.
The critical temperature is given by
$k_B T_c= 2\pi m(n/\zeta(3/2))^{2/3}$.

With all these building blocks modified in presence
of $\vc{k_s}$, we get the
results:
\begin{widetext}
\begin{eqnarray}
G_{11}(\vc{k},\omega)&=&
\frac{(\omega-\vc{k_s}.\vc{q}/m+\epsilon_{\vc{q}})
[1-2g\chi_{nn}^0(\vc{q},\omega)]+gn_c[1+
2g\chi_{nn}^0(\vc{q},\omega)]}
{\Delta(\vc{q},\omega)}
\\
G_{12}(\vc{k},\omega)&=&-\frac{gn_c[1+2g\chi_{nn}^0(\vc{q},\omega)]}
{\Delta(\vc{q},\omega)}
\\
\chi_{nn}(\vc{q},\omega)&=&\frac{
\left[\left(\omega-\vc{k_s}.\vc{q}/m \right)^2
-\epsilon_{\vc{q}}^2\right]
\chi_{nn}^0(\vc{q},\omega)+
2n_c \epsilon_{\vc{q}}[1+g\chi_{nn}^0(\vc{q},\omega)]}
{\Delta(\vc{q},\omega)}
\\
{\Delta(\vc{q},\omega)}&=&
\left[\left(\omega-\vc{k_s}.\vc{q}/m \right)^2
-\epsilon_{\vc{q}}^2\right]
[1-2g\chi_{nn}^0(\vc{q},\omega)]-2gn_c \epsilon_{\vc{q}}
[1+2g\chi_{nn}^0(\vc{q},\omega)]
\end{eqnarray}
\end{widetext}

\begin{figure}[htb]
\centerline{\includegraphics[width=5.0cm]{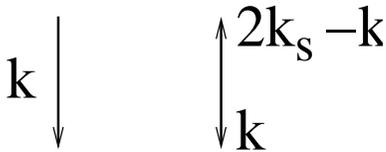}}
\caption{Representation of the normal and anomalous Green's 
functions in presence of a condensate velocity $\vc{k_s}$}
\label{fig0}
\end{figure}

\begin{figure}[htb]
\centerline{\includegraphics[width=5.0cm]{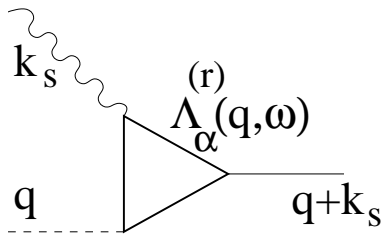}}
\caption{Representation of the regular vertex part 
$\Lambda_\alpha^{(r)}$ 
in presence of a condensate velocity $\vc{k_s}$ coupling 
a particle line of momentum $\vc{q}+\vc{k_s}$ with a condensate line 
with momentum $\vc{k_s}$ and an interaction line with momentum 
$\vc{q}$. }
\label{fig00}
\end{figure}


We are looking for the zeroes of $\Delta(\vc{q},\omega)$ 
the dielectric function. 
The real
part and imaginary part of the roots $\omega$ correspond to
the collective excitation and the Landau
damping respectively and have been determined previously in the case
of thermodynamic equilibrium
\cite{condenson,Fliesser}.
For low momentum excitations, 
we consider the limit $\omega \rightarrow 0$ and $|\vc{q}| \rightarrow 0$ 
and introduce the complex velocity of sound by $\omega / |\vc{q}|=c$. 
The only difference is that now we must consider terms of the form 
$\vc{k_s}.\vc{q}/|\vc{q}|$. If the wavevector $\vc{q}$ 
is parallel to the critical
velocity, we finnd the following generalization:
\begin{eqnarray}\label{lowdisp}
\frac{(c-v_s)^2-c_B^2}{(c-v_s)^2+c_B^2}=
\frac{2 g}{k_B T \lambda_{th}^3}
\chi_n (\frac{c}{(2k_B T/m)^{1/2}}) 
\end{eqnarray}
where
\begin{eqnarray}\label{chis}
\chi_n (s)= 
\frac{1}{\sqrt{\pi}}\int_{-\infty}^\infty
dt \frac{t}{s-t} \frac{1}{e^{t^2-\beta(\mu-2gn)}-1}
\end{eqnarray} 
is the dimensionless response function 
defined in the upper half of the complex plane. 
{\it The critical velocity is defined as the velocity beyond which 
the amplitude of the collective excitations 
grows exponentially}. In other words, if the imaginary part of any solution $c$ 
changes its sign from negative to positive, then a transition from metastability to instability occurs. 
The point at which this transition occurs corresponds to the situation 
where ${\rm Im}c=0$ in Eq.(\ref{lowdisp})  
and, consequently,  the imaginary part in the r.h.s. of (\ref{lowdisp}) must 
be equal to zero. This condition is satisfied provided that the imaginary 
part of Eq.(\ref{chis}) is zero which is the case only for ${\rm Re}c=0$. 
Thus we are left with a closed equation for the critical velocity 
$v_s$ that can be solved exactly:
\begin{eqnarray}
v_s&=&c_B\sqrt{\frac{k_B T \lambda_{th}^3 +2g\chi_n (0)}
{k_B T \lambda_{th}^3 -2g\chi_n (0)}} \\
&=&
c_B\sqrt{\frac{k_B T \lambda_{th}^3 -2 g\, g_{1/2}
(e^{\beta(\epsilon_\vc{k_s}-gn_c)})}
{k_B T \lambda_{th}^3 +2g \,g_{1/2}(e^{\beta(\epsilon_\vc{k_s}-gn_c)})}}
\end{eqnarray}
It is common in the literature ( e.g. \cite{textbook}) to determine 
the critical velocity from the {\it necessary } condition 
that the release energy of an excitation from the moving condensate 
becomes positive. In contrast, we use  the stability condition 
on the dynamics of the collective excitation which is {\it necessary and 
sufficient} to determine the critical 
velocity and this requires the  knowledge
of the Landau damping. As a consequence, this different criterion 
is not based on the energetic comparison between two thermodynamic 
states alone but rather on dynamical notions. 

The critical velocity is calculated as a function of the temperature using 
(\ref{state}). The 
dependences are displayed in Fig.3 and 4 for $gn/k_B T_c=0.1$ and 
$0.3$ respectively. In both figures, 
the critical velocity is lower than the sound 
velocity for temperatures not too close to the transition point 
by an amount which can exceed 10 percents. This difference 
is due to the particular dependence of the self-energy matrix 
on the momentum and the frequency in the presence of depletion. 
This dependence is needed 
in order to calculate the sound velocity whereas only  
the static expressions are required for the critical velocity. 
For comparison, we have also plotted the sound velocity obtained 
in the Popov approximation $c_P=\sqrt{gn_c/m}$ \cite{Popov}. 
That approximation is gapless
but not conserving and 
the self-energy matrix does not depend on $\vc{k}$ 
and $\omega$. Knowing that $\Sigma_{11}=2gn$ and $\Sigma_{12}=gn_c$, 
the application of the 
previous reasoning to this approximation allows to deduce that the 
spectrum of collective excitations does not contain any imaginary 
part. Since no dynamical instability appears, 
only the condition 
of sign change in the energy spectrum 
allows to determine  
the critical velocity which in this case corresponds to the sound velocity.  
For $T/T_c > 0.8$, we notice 
about a factor 2 of difference between the critical values obtained 
in the two approaches. As a consequence, the requirement that 
an approximation 
should be conserving modifies drastically the prediction.

\begin{figure}[htb]
\psfrag{t}[c]{$T/T_c$}
\psfrag{v}[b]{\rotatebox{270}{$v_s$}}
\centerline{\includegraphics[width=7.0cm]{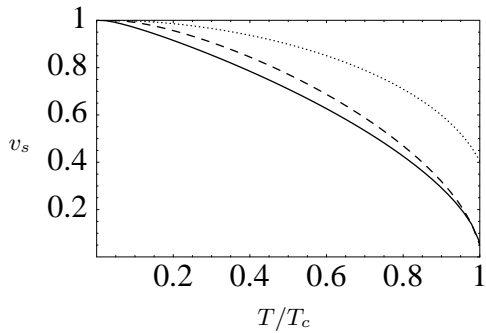}}
\caption{Critical velocity (full line) in units of $\sqrt{gn/m}$ 
as a function of the temperature for $gn/k_B T_c=0.1$.  
The sound velocity (dashed line) and the Popov approximation $c_B$ 
(dotted line) are represented 
for comparison.}
\label{fig1}
\end{figure}

\begin{figure}[htb]
\psfrag{t}[c]{$T/T_c$}
\psfrag{v}[b]{\rotatebox{270}{$v_s$}}
\centerline{\includegraphics[width=7.0cm]{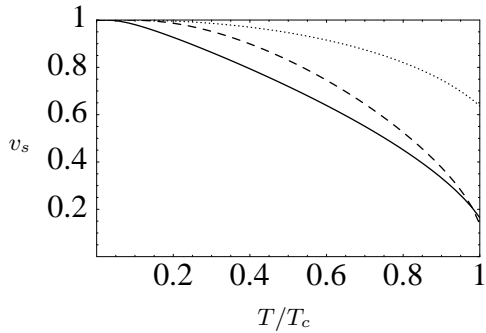}}
\caption{Same as for Fig.\ref{fig1} but for $gn/k_B T_c=0.3$}
\label{fig2}
\end{figure}

The influence of the relative velocity on the condensate population 
is also analyzed. Since the chemical potential depends on 
the relative velocity, the equation of state (\ref{state}) shows that 
the condensate fraction decreases when the 
velocity increases. This result is plotted in Fig.5 where we compare 
the two situations without relative velocity and at critical 
velocity which show only a minor difference of about few percents. 

\begin{figure}[htb]
\psfrag{t}[c]{$T/T_c$}
\psfrag{v}[b]{\rotatebox{270}{$\displaystyle \frac{n_c}{n}$}}
\centerline{\includegraphics[width=7.0cm]{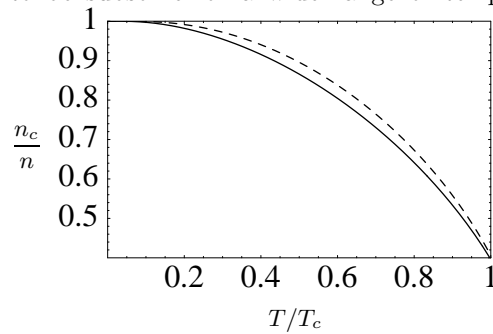}}
\caption{Condensate fraction as a function of temperature for $gn/k_B T_c=0.3$ 
at the critical velocity (full line) and in the absence of a relative 
velocity (dashed line). }
\label{fig5}
\end{figure}


In conclusion, we determine the critical velocity from the dynamics of the collective 
excitations using the gapless and conserving 
generalized random phase 
approximation. The calculated value displays relevant differences 
in comparison with the Popov approach and appears to be subsonic for a wide range of 
temperature indicating that the presence of thermal excitations 
facilitates the creation of an instability with the condensate. 

\centerline{\bf ACKNOWLEDGMENTS}
PN aknowledges support from the german AvH foundation.


\end{document}